\documentclass{iopconfser}
\usepackage{caption}
\usepackage{graphicx}
\usepackage{amsmath}
\usepackage{subfig}
\usepackage{multirow}
\usepackage[toc,page]{appendix}
\usepackage{hyperref}
\usepackage{listings}
\usepackage{xcolor}
\usepackage{framed}
\usepackage{array}
\usepackage{geometry}
\usepackage{float}
\usepackage{amssymb}
\usepackage[absolute,overlay]{textpos}
\begin{document}

\title{Are there critical aspects in the time, energy and angular distributions of SN1987A?}

\author{Veronica Oliviero$^1$, Riccardo Maria Bozza$^1$, Vigilante Di Risi$^1$, Giuseppe Matteucci$^1$, Giulia Ricciardi$^1,^2$ Francesco Vissani$^3$ }
\affil{$^1$Dipartimento di Fisica E. Pancini, Università degli Studi di Napoli Federico II}
\affil{$^2$Istituto Nazionale di Fisica Nucleare, Sezione Napoli}
\affil{$^3$Laboratori Nazionali del Gran Sasso (INFN), Assergi (AQ)}

\email{veronica.oliviero2@unina.it}

\begin{abstract}
Supernova neutrinos are of considerable importance for ongoing research in astrophysics, nuclear and particle physics. Existing simulations of this complex event are increasingly sophisticated, but the accuracy with which they describe the emission is unknown. The only event observed so far with neutrino telescopes, SN1987A, still plays a crucial role and deserves to be studied meticulously. With this in mind, we have undertaken a refined analysis of the observations, taking into account the knowledge gained over the past decades. In this work, we consider a new parameterised model of electron antineutrino emission and test its adequacy in describing the observed distributions of energy, time and angle. The values of the model parameters derived from the data and their uncertainty intervals are presented and their significance is discussed.
\end{abstract}

\section{The new model}
The simplest approximation of the instantaneous neutrino spectrum emitted by a supernova, which is justified on physical grounds, is the thermal distribution (which in practice consists of a black-body model). However, all simulations show that the initial emission is much more intense, indicating that some physical aspects are missing and a more elaborate model is needed.
A new model \cite{sym13101851}, which is an upgrade of the one proposed in \cite{PAGLIAROLI2009163} that at the time enhanced the one in \cite{PhysRevD.65.063002}, related to SN1987A neutrino observations, has been proposed.
This new model proposes a parametrization that includes two emission phases, which are distinct both physically and mathematically: accretion and cooling.
While the cooling phase gives a thermal emission due to the black body surface, the accretion is a volume emission. 
\newpage
It is taken into account due to reactions occurring in the initial moments following the onset of gravitational collapse. Specifically, for electronic antineutrinos:
\begin{equation}
n + e^+ \rightarrow p + \nu_e. 
\end{equation}
This phase occurs in a neutron-rich region just above the nascent star, where matter accretion is at its peak and the temperature is sufficiently high to produce a thermal population of positrons. It is important to note that this second component is inherently non-thermal.
\subsection{Cooling phase}
Looking at the thermal cooling , we have a rate of antineutrinos emission equal to:
\begin{equation}
\dot{N}_{\nu,c}(E_{\nu}, t) = \frac{c}{(hc)^3} \times \pi R^2_{ns} \times \frac{4 \pi E_{\nu}^2}{1 + \exp(E_{\nu}/T_c(t))} \, dE_{\nu}
\end{equation}
In this approximation, the average neutrino energy is determined by the well-known Fermi-Dirac formula: \( E_{\nu,c} = 3.15 \, T_c \). The luminosity follows the black-body law:
\begin{equation}
L_c \propto R_{ns}^2 \, T_c^4.
\end{equation}
The flux depends on these parameters: the radius of the neutron star \( R_{ns} \) and its temperature \( T_c \), which typically vary over time. When fitting observations with this spectrum, these parameters serve to quantify both the emission intensity and the average energy of the neutrinos emitted.

\subsection{Accretion phase}
After the bounce, the shock wave propagating into the outer core of the star, looses energy and eventually gets stalled. It forms an accreting shock that encloses a region of dissociated matter and hot \(e^+\)\(e^-\) plasma. Neutrons in the nearby space become the target for the flux of thermal positron giving rise to neutrino emission. \\
If we call \(N_n\) the number of neutrons participating in the process and \(T_a\) the temperature of the positrons, the rate of emission of antineutrinos will be given by:
\begin{equation}
\dot{N}_{\nu,a}(E_{\nu}, t) = \frac{c}{(hc)^3} \times N_n(t) \times \sigma_{ne}(E_{\nu}) \times \frac{g_e \times 4 \pi E_e^2}{1 + \exp(E_e/T_a(t)} \, dE_{\nu}
\label{a}
\end{equation}

where \(g_e = 2\) is the spin factor, while \(E_e(E_v)\) and \(\sigma_{ne}(E_v)\) are as in \cite{PAGLIAROLI2009163}. 
As we can see from Equation \ref{a}, the accretion component depends on two parameters \(N_n\) and \(T_a\), both are time dependent. Regarding \(N_n\) parameter it can be expressed as follow:
\begin{equation}
    N_n = \xi_n \times \frac{M_\odot}{m_n}
\end{equation}
where \(\xi_n\) is the fraction of outer core mass that is composed by neutron and exposed to positron flux.
The neutrons around the newborn neutron star are the target of positrons. Therefore, their number is constrained on a physical basis to be less than a fraction of the solar mass. This aspect has been critically discussed in \cite{PAGLIAROLI2009163} and especially in \cite{PhysRevD.65.063002}, whose analyses suggest relatively large values of the number of neutrons, on the edge of the acceptable region. From the literature, we expect a value of \(\xi_n < 0.4\).
We will take this aspect into account later testing the model.
\subsection{Time dependence of the model}
In accordance with the general framework previously outlined, our model consists of two distinct components of the flux. The first component is the accretion phase, which lasts less than a second, and the second is the cooling phase, which is much less intense but of a longer duration. We denote these two time scales as \( \tau_a \) and \( \tau_c \), where \( \tau_a \) is less than 1 second and \( \tau_c \) is approximately several seconds. To achieve a time parameterization of the neutrino emission, we have described the total luminosity, as detailed in \cite{sym13101851}.
Following this modelization we obtain:\\
\begin{minipage}{0.45\textwidth}
\begin{equation}
T_a(t) = 0.6 \times T_0
\label{T}
\end{equation}
\begin{equation}
\xi_n(t) = \xi_{n0} \times \textit{F}(t, t_{max}, \tau_a, 2, n_a) \quad 
\label{xi}
\end{equation}
\end{minipage}
\begin{minipage}{0.45\textwidth}
\begin{equation}
T_c(t) = T_0 \times ^4\sqrt{\textit{F}(t, t_{max}, \tau_c, 1, n_c)}
\label{Tc}
\end{equation}
\begin{equation}
R_{ns}(t) = R_{ns0}\quad
\label{R}
\end{equation}
\end{minipage}\\

In the Equation \ref{R} we have $R_{ns}$ that is radius of the neutron star. Instead, in the Equation \ref{T} and \ref{Tc} we have \(T_0\) that is the temperature at \(t=t_{max}\), which is the time of maximum luminosity. This is true because when the time is equal to \(t_{max}\) the positive function \(\textit{F}(t,t_{max},\tau_c, \alpha, n)= 1\).
As we can see in  \cite{sym13101851}, the function \textit{F} has several fundamental properties, like for example: it's positive and has a maximum, at which it is equal to one. 
On the other hand, Equation \ref{T} is crucial to connect consistently the accretion phase to the cooling one.
Our model consistently follows all simulations to date, which do not include a point of discontinuity between the two phases.
However at the moment, all we know about the accretion phase is that it is predominant at the beginning, as the neutron star exists almost immediately.
What we have done to match the two phases with continuity is to put their respective average energies equal, and this is verified when the Equation \ref{T} holds.
We can say that our parameterisation gives a very accurate description on time while on energy it is more basic.
We chose this parameterisation to verify the existence of the accretion phase, but it is not designed to describe the details, which are not visible with SN1987A.
Still, care is taken to remove the questionable feature of previous parameterisations, for which the maximum of flux occurs exactly for t=0.
Summarizing all, our model has six parameters that describes the two emission phases:
\begin{equation}
 \{ T_0 , \xi_{n0} , R_{ns0} , t_{max} , \tau_a ,  \tau_c \}  
\end{equation}
In the next section we will discuss about how we tested the model giving benchmark values to the parameters.

\section{Goodness of fit}
To be be sure that our model is consistent with the observation we need to verify the goodness of fit. We choose to use the Cramér–von Mises criterion.
The Cramér-von Mises test is a goodness-of-fit test that compares the empirical cumulative distribution function (ECDF) of a sample to the cumulative distribution function (CDF) of a specified theoretical distribution. The goal is to determine if the data follow the theoretical model.
The test statistic \( w^2 \) is computed as:
\begin{equation}
 w^2 = n \int_{-\infty}^{\infty} \left[ F_n(x) - F(x) \right]^2 \, dF(x)   
\end{equation}
where: \( n \) is the sample size, \( F_n(x) \) is the ECDF and \( F(x) \) is the theoretical CDF.\\
The first step was give benchmarks values to the parameters, in accordance with the literature \cite{sym13101851}:
\begin{equation}
   { T_0 = 4.2 \, \text{MeV},\ \xi_{n0} = 0.02,\ R_{ns0} = 12 \, \text{km},\ t_{max} = 0.1 \, \text{s},\ \tau_a = 0.3 \, \text{s},\ \tau_c = 5 \, \text{s} }
\end{equation}
After, the Cramer Test was performed using the SN87A neutrino events observed by the experiments: Kamiokande-II, IMB and Baksan \cite{PhysRevLett.58.1490, PhysRevLett.58.1494,ALEXEYEV1988209}. The detection of all the three experiments comprises a total of 29 events in a time span of about 30 seconds.
We calculate the interaction rates in the Kamiokande-II, Baksan and
IMB detectors and quantify the goodness-of-fit of our model.

\subsection{Results with SN87A data}
\begin{figure}[htpb]
    \centering
	\includegraphics[width=0.4\textwidth]{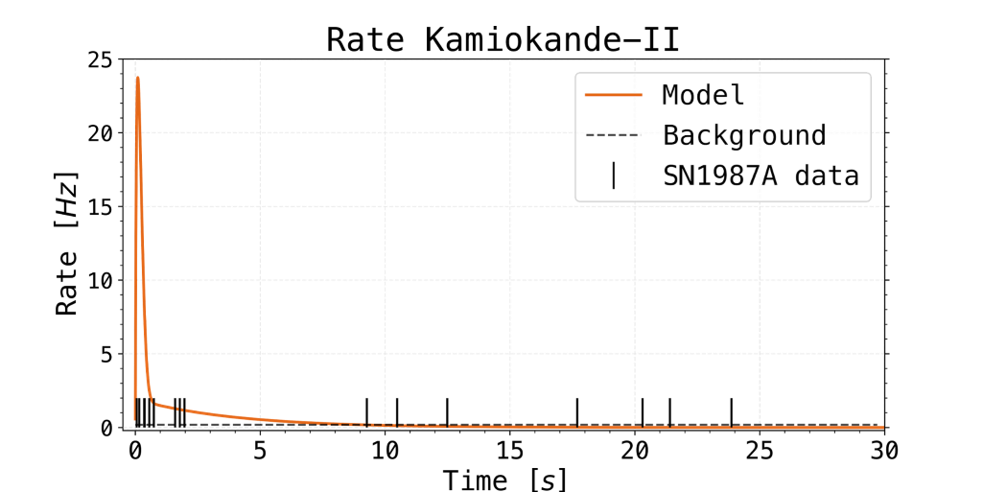} 
	\qquad
	\includegraphics[width=0.4\textwidth]{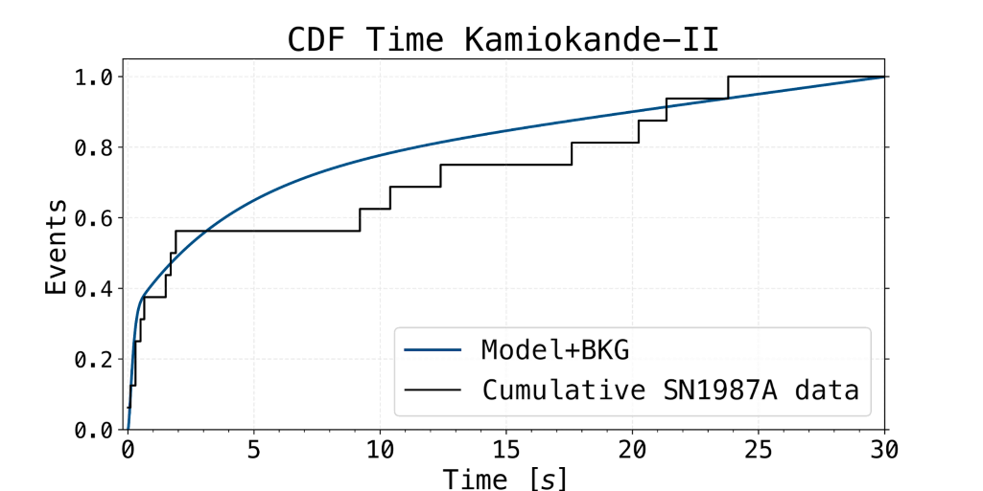} 
	\qquad
	\caption{\footnotesize\centering\emph{On the left: Differential time distribution of the events from SN1987A, as observed in Kamiokande-II (black vertical lines) compared with the differential counting rate predicted in the model (red curve) for the value of the parameters indicated in the text. On the right: Cumulative distribution functions (CDF) for the same flux, regarded as a model of SN1987A emission.}}
	\label{timek2}
\end{figure}

\begin{figure}[htpb]
    \centering
	\includegraphics[width=0.4\textwidth]{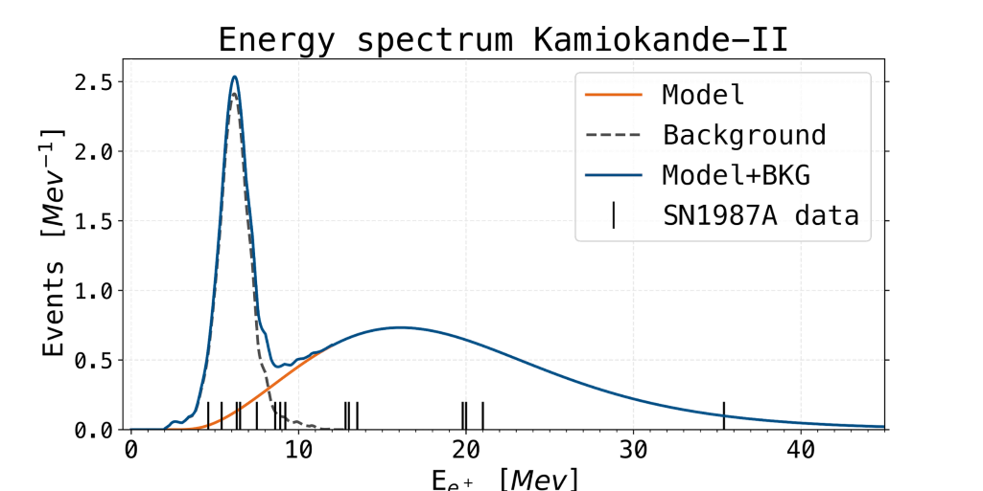} 
	\qquad
	\includegraphics[width=0.4\textwidth]{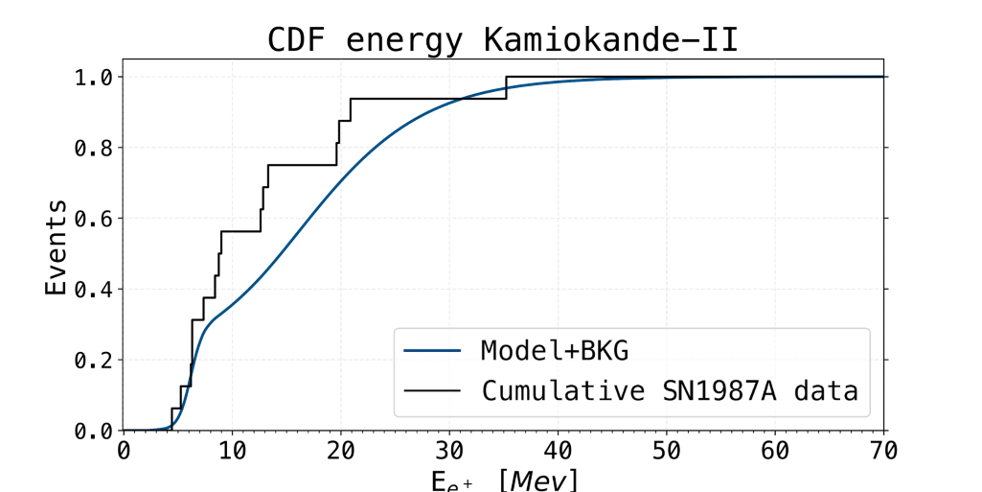} 
	\qquad
	\caption{\footnotesize\centering\emph{On the left: Differential energy distribution of the events from SN1987A, as observed in Kamiokande-II (black vertical lines) compared with the differential counting rate predicted in the model (red curve) and signal + background (blue curve) for the value of the parameters indicated in the text. On the right: Cumulative distribution functions (CDF) for the same flux, regarded as a model of SN1987A emission.}}
	\label{energyk2}
\end{figure}

\begin{figure}[htpb]
    \centering
	\includegraphics[width=0.4\textwidth]{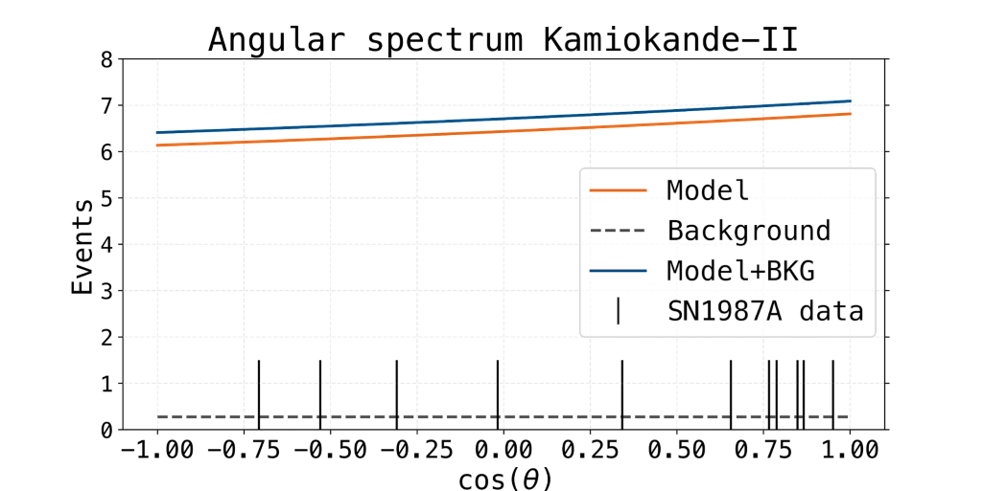} 
	\qquad
	\includegraphics[width=0.4\textwidth]{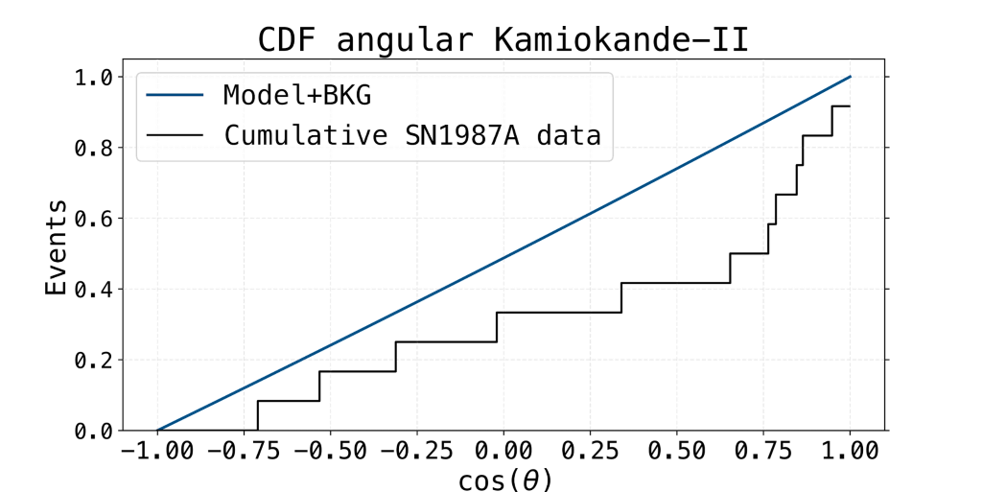} 
	\qquad
	\caption{\footnotesize\centering\emph{On the left: Differential angular distribution of the events from SN1987A, as observed in Kamiokande-II (black vertical lines) compared with the differential counting rate predicted in the model (red curve) and signal + background(blue curve) for the value of the parameters indicated in the text. On the right: Cumulative distribution functions (CDF) for the same flux, regarded as a model of SN1987A emission.}}
	\label{anglek2}
\end{figure}

 We present here the distribution functions of time, energy and angle. The study was done on all three experiments in the same way, but here we only show for Kamiokande-II all the distributions, while for IMB and Baksan we report only the results of the test.
 We can see in Figure \ref{timek2} for the time distribution, events accumulate rapidly at the beginning, aligning with the concept of an initial phase of emission due to accretion. In Figure \ref{energyk2} for the energy distribution, there is a distinct group of low-energy events (including the last four events, which are often excluded in SN1987A data analyses) that can be attributed to background processes. Instead in Figure \ref{anglek2} for the angular distribution we can see that the agreement between data and model is not so good as that for time and energy but still acceptable.
 The data seem to prefer forward scattering \cite{Costantini:2004ry}, a property that is not expected in our isotropic angular distribution spectrum. It is premature to determine whether this feature is a statistical fluctuation, a systematic property of the detector, or a physical insight into the phenomenon. Nonetheless this topic is of significant interest, especially considering that IMB also exhibit the same pattern (Baksan did not reconstruct the angles). \cite{Vissani:2008wbj}

\begin{table}[H]
\centering
\begin{tabular}{|>{\raggedright\arraybackslash}p{3cm}|>{\raggedright\arraybackslash}p{3cm}|>{\raggedright\arraybackslash}p{3cm}|>{\raggedright\arraybackslash}p{3cm}|}
\hline
\textbf{p-values} & \textbf{Kamiokande-II} & \textbf{Baksan} & \textbf{IMB} \\ \hline
\textbf{Rate} & Cramer: 46\% & Cramer: 83\% & Cramer: 44\% \\ \hline
\textbf{Energy} & Cramer: 17\% & Cramer: 55\% & Cramer: 17\% \\ \hline
\textbf{Angle} & Cramer: 8\% & N/A & Cramer: 9\% \\ \hline
\end{tabular}
\caption{P-values from the three experiments using Cramér's test. Smirnov's test was also perfomed to estimate g.o.f, as the results are very similar we only report those of Cramer's test.}
\label{table:1}
\end{table}
Furthermore, looking at the results of the Cramer test shown in Table \ref{table:1}, we note that the p-values obtained show that for all three experiments the theoretical time distribution is perfectly consistent with the observations. The same consideration can also be made for the energy distributions. whereas for the angular ones, it is a little different: Baksan does not reconstruct the angles, so we have no information; for IMB and Kamioikande-II, on the other hand, the agreement between data and model is not excellent but still acceptable.

 In conclusion, this model is consistent with the observations from SN1987A, even before optimizing the parameter values.

\section{Discussion and Outlook}
In conclusion we can say that the agreement with the temporal and energy distributions of the model is excellent; that with the angular distributions not excellent, but acceptable. The results have been validated with previous literature, where possible.
We conclude that our model can effectively describe an anti-neutrino burst from a core-collapse supernova and
provide useful insights into the astrophysics of emission and gravitational collapse.
As the next steps for our analysis, we will do a best fit analysis to actually estimate the parameters of our model.
We choose to use the likelihood maximization function, both for the single experiment and by combining all experiments. 
At the moment, We are consolidating the best-fit results and are planning to proceed with estimating the confidence intervals for our parameters.

\newpage
\bibliographystyle{unsrt}
\bibliography{biblio}

\end{document}